\documentclass[twopage,11pt] {article}  
\usepackage{graphicx}
\begin{document}           

\title{
\begin{flushright}
\small{FZU-D 20040565 \\[5mm]}
\end{flushright}
Quantum theory of microworld and the reality}  
\author{Milo\v{s} V. Lokaj\'{\i}\v{c}ek \\
   Institute of Physics, AVCR, Na Slovance 2, 18221 Prague 8, Czech Republic } 
\date{December 2004}

\maketitle                 

\pagestyle{myheadings} \thispagestyle{plain} \markboth{Milo\v{s}
V. Lokaj\'{\i}\v{c}ek}{Quantum theory of microworld and the
reality} \setcounter{page}{1}

 {\bf  Abstract }

The discoveries in the end of the $19^{th}$ century and in the
beginning of the $20^{th}$ century have paved the way for the
physical science of the new era. It was shown that the hitherto
basic stones (atoms) of the matter world were not quite stable and
did not represent the smallest matter objects. And especially,
that the energy of the light was not transmitted continuously but
in small quanta.

The mathematical model of orthodox quantum mechanics was
accepted by the physical community as the only theory of
microworld, even if it involved a series of strange (paradoxical)
characteristics. It is possible to say that the things started to
change in the end of the $20^{th}$ century. More attention has
been devoted to the problem of time reversibility and to the
question whether it is not possible for the microworld to exhibit
irreversible evolution similar to that of macroscopic world.

To understand better these physical problems and the contemporary
theoretical attempts it is necessary to follow at least shortly
the most important points of the whole physical story since the
end of the $19^{th}$ century. It will be shown that one of the
basic quantum-mechanical assumptions (physical interpretation of
the general mathematical superposition principle) has been
introduced without any actual reason and experimental tests;  some
misleading arguments having been used, in addition to.

The difference between microscopic and macroscopic worlds
diminishes significantly if the quantum-physical model starts
fully from time-dependent Schr\"{o}dinger equation and the general
superposition principle linking  states with different physical
properties is abandoned; divers physical meanings of different initial conditions of
individual solutions being fully respected. Such a model
(including suitably extended Hilbert space) may be regarded as a
generalization (or even as a mere quantum adaptation) of the
classical approach. The physical systems with both fixed and
changing numbers of objects may be now described in the framework
of one common mathematical model.

 \section{ Introduction  }
 \label{secI}
It is possible to say that the time flow (time evolution) has
represented important problem in quantum mechanics practically
during the whole past century. We shall start, therefore, in this
introduction with a brief survey of the beginnings of quantum
mechanics. This beginning is linked closely with the name of E.
Schr\"{o}dinger  \cite{schr}, who believed firmly in wave nature
of matter. He proposed his famous wave equation in 1925 and showed
that it was possible to reproduce all main results obtained
earlier on the basis of Hamilton equations. All characteristics of
a physical system could be then derived with the help of wave
function $\Psi({\mathbf r},t)$ where ${\mathbf r}$ represented
coordinates of all matter objects. A great success has been seen
in that in the case of bound systems the discrete atom energy
levels have corresponded to eigenvalues of corresponding
Hamiltonian \cite{schr2}. The wave function $\Psi$ was interpreted
as probability distribution by M. Born \cite{born}. N. Bohr (see
\cite{bohr28}) attributed then the probabilistic properties
directly to individual matter particles; Heisenberg's \cite{hei}
uncertainty relations being linked with them.

 A further step was done by J. von Neumann \cite{vonne} who showed that
individual $\Psi$-functions at given $t$ values may be represented
by vectors in the Hilbert space spanned on eigenfunctions of the
corresponding Hamiltonian (determining the total energy of a given
physical system). The additional assumption has been then introduced into the
corresponding model: general mathematical vector superositions have 
been interpreted as physical relations between divers physical states. 
And Pauli's criticism formulated already in
1933 has concerned just it. Pauli \cite{pauli} has showed that
under the mentioned conditions a correct representation of time
evolution (introduction of the time operator $T$ in the Hilbert
space) has required for the Hamiltonian to possess continuous
spectrum belonging to the whole real interval
$({-\infty,+\infty})$, which contradicts the necessity of energy
being positive (or at least limited from below). 
It contradicts in principle also the chracteristics following from the 
corresponding time-dependent Schr\"{o}dinger equation.

Pauli's problem has been, however,  related later to another one
concerning the non-unitarity of exponential phase operator
        \[ \mathcal{E} \;\;= \;\; e^{-i\Phi}\;\;  \]
where the phase $\Phi=\omega T$ is proportional to time. The exponential
phase operator for linear harmonic oscillator was defined by Dirac
\cite{dirac} already in 1927 and it was assumed to be unitary. But
it was shown later by Suskind and Glogower \cite{sus} that the
operator $\mathcal{E}$ defined in such a way was isometric only
(as it will be shown in Sec.~\ref{ssecIII13}); see also the review
by Lynch \cite{lynch}. This fact and Pauli's critique have been
regarded all the time as following from one common source.
However, both these problems will be shown to be solved only if
they are handled as two different problems.

In Part~\ref{secIV} we will present not only the solution of both the
problems but also that the new quantum-theoretical model may be brought to
harmony with available experiments (including those having been performed in
the last ten years). However, to provide necessary insight into the related
problems the contemporary status of corresponding physical knowledge will
be summarized, first. In Part~\ref{secII} we shall discuss the state of physical
research in the end of 19$^{th}$ century devoting main attention to problems
preparing the era of modern physics. The problems concerning the contemporary 
quantum physics will be then explained in Part~\ref{secIII}; with the respect
to points being solved in Part~\ref{secIV}. Final summary will be given 
in Part~\ref{secV}.

  \section {Physics in the end of the $19^{th}$ century   }
 \label{secII}
 \subsection{Light waves and matter objects  }
 \label{ssecII1}

One can say that till the end of the $19^{th}$ century two
different worlds existed one beside another: matter objects and
the light or electromagnetic waves. The behavior of the formers
was ruled by the laws that were formulated originally by G.
Galilei and I. Newton. In the $19^{th}$ century they were then
described systematically, e.g., with the help of Hamilton
equations; W. Hamilton (1805-65).

 The other part (light and electromagnetic waves) of the observable
world was governed by equations of J. Maxwell (1831-79) who
unified the description of electric and magnetic phenomena in 1865
\cite{maxw}. The relations of electromagnetic waves to matter
world were given by the electric and magnetic properties of matter
objects. These relations were studied on the basis of different
frequency and temperature dependencies, especially with the help
of the so called black body radiation.

In both the cases the nature phenomena were described with the
help of differential equations that provided fully causal and
practically deterministic behavior of all nature processes. The
evolution of any physical system was (in principle) uniquely
determined when its state was known fully at one time instant.

 \subsection{Atoms and molecules  }
 \label{ssecII2}
The assumption that matter objects consisted from some not further
divisible particles was formulated already by Democritos {\mbox
(460-370 BC)} in the old Greece. In the modern age it was
formulated again by P. Gassendi (1592-1655) and R. Boyle
(1627-1691); and later developed by M. V. Lomonosov (1711-65) and
J. Dalton (1766-1844); e.g., the atom weight was introduced and
the first table of atom weights proposed by Dalton in 1803. D. I.
Mendeleev (1834-1907) compounded then the periodical law of
elements in 1869.

Besides the atoms the existence of molecules consisting of two or
more atoms were assumed, too. All the conclusions were derived
from the results observed with the help of chemical reactions and
corresponding laws. Similar conclusions could be drawn from a
great amount of various data. And further predicted properties
were found to agree with observations.
 Thus, the existence of atoms was accepted by the whole physical
community even if the atoms and molecules could not be observed
directly at that time.

  \subsection{Thermodynamics of gas systems  }
 \label{ssecII3}

The grounds of heat theory on the basis of kinetic energy of atoms
and molecules were formulated cca in 1790. The average kinetic
energy of atoms (or molecules) was assumed to rise proportionally
to the measured temperature.

The molecular theory was taken also as the basis of thermodynamics
of gas systems.
 The basic behavior of such a system consisting of many molecules has
been described by the state equation
      \[  \frac{pV}{T} \;=\; const  \]
where $V$ is volume, $p$ - pressure, and $T$ - temperature. The
given equation holds when the gas system is in equilibrium. The
system has its inner energy $U$ that rises with temperature $T$:
     \[ dU \;=\; \beta dT . \]

Entropy $S$ of the system has been then defined by
       \[  dS \;=\; \frac{dQ}{T}  \]
where $Q$ is heat content; its change being given by
       \[  dQ \;=\; dU \;+\;pdV .   \]

In an enclosed system the entropy change $dS$ is always positive,
the entropy rising to a maximum value when the system reaches its
equilibrium state.  L. Boltzmann (1844-1906) linked this
(deterministic) entropy rise with the tendency of the system to go
to uniform average probability distribution of individual 
molecules \cite{boltz}. In 1867 he proposed the equation
          \[   S \;=\; k \log w   \]
where  $w$ is the probability of molecule distribution determined
in the classical way.

The distribution probability rise was denoted as a basic natural
law, even if it is in principle the result  of collision and
diffusion processes (and corresponding rules). It is evident that
Boltzmann was influenced strongly by the positivistic philosophy
(refusing any ontology and metaphysics) as practically all
physicists at that time. Agreement of model predictions with
measured numerical values was for them quite sufficient to regard
corresponding model (hypothesis) as verified; without testing its
logical structure.

  \subsection{Black body radiation }
  \label{ssecII4}

As already mentioned the matter world consisted at that time of
two different media: the matter objects moving according to
equations of motion and the light (electromagnetic) waves being
transmitted by swinging of the ether present in all space.  The
interactions between the electromagnetic and light waves were
described by Maxwell's equation \cite{maxw}; waves being excited
by moving electrically charged particles.

The mutual interactions between heated objects and light were
studied with the help of the so-called black body radiation. The
frequency spectra at different temperatures inside a body (being
quite black at the absolute zero) were measured. In 1896 Wien
(1864-1928) showed that all experimental frequency spectra (for
higher values of $\nu$) may be described with the help of the
formula
  \[  \rho (\nu)\;=\; a\nu^3e^{-\frac{b\nu}{T}}  \]
where $\nu$ is the corresponding light frequency, and $a, b$ are
parameters adapted according to the material of the black body.

The theoretical formula derived by Rayleigh (1842-1919) and Jeans
(1877-1946) on the basis of continuous transfer of energy between
the two different objects (body and ether) was equal to
     \[  \rho (\nu)\;=\; \frac{8\pi}{c^3}\nu^2k_BT ,  \]
i.e. very different from the experimentally measured spectra.

This contradiction was solved by M. Planck (1858-1947) in 1900 by
assuming that the light energy was transferred always in finite
amounts and that only multiples of a basic energy amount might be
emitted \cite{plan}. We will describe his approach in
Sec.~\ref{ssecII6}.

Planck's quantum idea opened the main way to the physics of the
$20^{th}$ century. The other important way concerned the problem
of ether, which will not be discussed in this paper; the problem being
mentioned only shortly in the next section.

   \subsection{Light ether  }
  \label{ssecII5}

When the light was transmitted not only through the air but also
through different transparent bodies the question arose how the
ether was related to matter objects. Is it quite immobile in the
whole world space? or: Does it move together with moving matter
objects?

Several experimental studies were performed that led to different
conclusions. The experiment of Michelson and Morley (1887) led to
the conclusion that the ether should be closely linked to the
motion of the earth. A quite opposite result could be derived from
the measurement of star abberations; the phenomenon discovered by
Bradley in 1727. The ether should be linked to the sun as the
observed fixed stars exhibited small ellipses. And according to
Fizeau's experiment (1851) the ether should be partially took
along by moving objects when the light velocity changed with the
velocity of water.

These three experiments put so different requirements on the ether
that its existence should have been doubted. That led Einstein
\cite{ein5} to refuse the existence of ether and to formulate the
theory of special relativity in 1905. He assumed that the light
velocity depended neither on the velocity of a moving system nor
on its actual direction.

We will not discuss this problem further here. We have mentioned
the theory of relativity as it belonged to two basic theories of the past
century applied to the description of the microscopic world, even
if these two theories have provided quite different pictures:
special relativity being deterministic as classical physics and
quantum mechanics refusing any causality. 

    \subsection{New discoveries at the break of centuries }
 \label{ssecII6}
The way to the new physics was opened mainly by discoveries in the
last five years of the $19^{th}$ century. They were: The X
radiation emitted from cathode tubes and discovered by W. Roentgen
(1895), ionizing radiation emitted by some heavy substances and
found by A. S. Becquerel (1896), and the discovery of electron by
J. J. Thomson (1897). These discoveries showed convincingly that
there is a very rich physics under the level of our direct
observations, which laid quite new claims to measurement
techniques.
 These new discoveries influenced strongly not only the physical
research, but they evoked also many new application regions. E.g.,
quite new branches of medicine arose along with the new branches
of physics.

However, as already mentioned the physical thinking was
fundamentally influenced by M. Planck \cite{plan} who introduced
the quantum idea in solving the problem of black body radiation.
He assumed that the exchange of energy between an oscillator in
the body and an electromagnetic (light) wave might occur only in
multiples of a basic energy quantum, i.e., the exchanged energy
might equal
                     \[  E_n \;=\; nh\nu  \]
where $\nu$ was radiation frequency and $h$ was the famous
constant of Planck; $n=1,2,.. $.

He assumed further that the energy of oscillators increased with
temperature and one could write for the numbers of states with
higher energy
              \[ N_n \;=\; N_0 e^{-\frac{nh\nu}{k_BT}}  \]
where $k_BT$ was Boltzmann factor. The average energy per one
oscillator in the equilibrium was then
      \[ {\bar E}\;=\; \frac{E_t}{N_t} \;=\;
               \frac{\Sigma_m\, N_m mh\nu}{\Sigma_m\, N_m}  \]
where $E_t$ is total energy and $N_t$ - the number of all
eigenoscillations. It holds then
         \[ {\bar E}  \;=\; \frac{h\nu}{e^{\frac{h\nu}{k_Bt}} - 1} . \]
And if one puts this average energy into the formula of Rayleigh
and Jeans (see Sec.~\ref{ssecII4}) instead of $k_BT$ it is
possible to obtain the following final formula \cite{plan}
   \[ \rho(\nu) \;=\; \frac{8\pi}{c^3}\frac{h\nu^3}{e^{\frac{h\nu}{k_BT}} -1} . \]
The last formula passes for very large and very low frequencies
into one of the two formulas introduced in Sec.~\ref{ssecII4}.

It is possible to say that the discoveries and ideas mentioned in
this section formed the new basis from which the physics of
microscopic world developed. The problem of photoeffect (discussed
in the next section) has belonged to the first phenomena from this
series.

  \section{Physics of the $20^{th}$ century   }
 \label{secIII}

   \subsection{Photoeffect and the birth of duality}
 \label{ssecIII1}

Planck assumed that the electromagnetic radiation should have been
interpreted as waves and that the cause of discrete energy
emission was hidden in oscillating matter objects. There were,
however, other experiments that were not in full harmony with such
an assumption.

It was found that metals irradiated by the light of higher
frequency emitted some charged particles. In 1902-3  Lenard
\cite{lenar} and Ladenburg \cite{laden} showed that these
particles were electrons. However, the electrons were emitted only
if radiation frequency was greater than a minimal value $\nu_0$.
It was found, too, that the energy of emitted electrons did not
depend on light intensity, but only on light quality, which could
not be interpreted on the basis of classical wave ideas. It
followed from this fact that the light energy in the wave front
could not be distributed in a continuous and uniform way.

Einstein \cite{einph} tried to solve the problem in the following
way: He assumed that the energy of electromagnetic wave came in
discrete units of $h\nu$ and each of these units occupied a very
small volume only. Therefore, he assumed that a radiation (light)
quantum had the same properties as a matter particle. Einstein
wrote for the photoelectric process the following equation
        \[ \frac{1}{2}mv^2\;=\; e V \;=\; E \;=\; h\nu - \Phi  \]
where $e$, $m$, and $v$ are the electric charge, mass and velocity
of the knocked-out electrons; $V$ is the minimal potential for
which no emission occurs  (boundary of the photoeffect); and
$\Phi=h\nu_0$ is the binding energy of an electron which
characterizes the given material (metal).
 Einstein assumed also in 1904 that the ratio
         \[   \frac{eV+\Phi}{\nu} \;=\; h  \]
is independent of the material and of the intensity and
frequency of light \cite{einph}, being equal to the constant that
was introduced earlier by Planck.

Thus the duality appeared in physical story for the first time.
Particles (photons) were the carriers of energy that belonged to
the corresponding electromagnetic field, while the electromagnetic
waves around these particles represented other characteristics.
The particle structure was   necessary to understand the photoeffect, while the
waves enabled  to explain the interference and diffraction of
electromagnetic radiation. The energy of individual photons has
equaled $h\nu$.

However, practically any physicist did not believe in the quantum
idea at that time. The situation changed in 1914-16 when R. Millikan
(1868-1953) confirmed all predictions of Einstein experimentally. 
Einstein obtained then the Nobel price for theoretical
predictions of photon in 1922.

   \subsection{Atom nucleus  }
 \label{ssecIII2}

The discovery of electron and the existence of photoeffect changed
also our view on the indivisibility of atoms. J. J. Thomson
(1856-1940) proposed the new atom structure in 1903. He assumed that
the atom was a very tiny sphere the electrons floated in.

A quite new result was then obtained by E. Rutherford (1871-1937)
in 1911 \cite{ruth}. He irradiated a tiny gold foil by a beam of
$\alpha$ particles. The scattering characteristics showed that the
atom mass was not distributed homogeneously over the whole
atom volume but concentrated in a much smaller volume.

It was the basis for atoms to be regarded as small planetary
systems: A positively charged heavy nucleus in the center and
negative electrons running around. The nucleus was thousand times
smaller than the whole atom. There were, however, problems with
such a model as according to electromagnetic theory (classical
physics) the corresponding systems could not be stable. The
electrons running round a nucleus had to emit electromagnetic
waves and could not move in stable orbits.

   \subsection{Atom model  }
 \label{ssecIII3}

The problem of a stable atom was solved by N. Bohr (1885-1962)
with the help of two additional postulates in 1913. He combined
the planetary model with the quantum picture of Planck and
Einstein. The two postulates were \cite{bohr3}:

1)  The postulate of stationary states. According to classical
physics the electrons might circulate in any distance from the
nucleus. However, Bohr assumed that only certain tracks with
discrete energies were possible and that the values of these
energies were
   \[ E_n \;=\; nh \;=\; \oint p\,dq \;, \;\;\; n\;=\; 1,2,3,...  \]
where  $p$ is the momentum of electron, $q$ - its distance from
the nucleus, and $n$ - the so-called main quantum number.

2)  The frequency postulate.  If the electron passed from one
stationary track with energy $E_n$ to the other one with energy
$E_m \,<\, E_n$,  the atom emitted spontaneously the energy
        \[ h\nu_{nm} \;=\; E_n - E_m  \,.  \]
Bohr obtained for these energies:
       \[  E_n \;\simeq \; -\frac{2\pi^2m_ee^4}{h^2n^2}   \]
where  $m_e$ is electron mass, and $e$ - its charge. The
frequencies of emitted light by hydrogen atoms corresponded to
frequency values that were measured by Balmer (1825-98) already in
1880.

N. Bohr assumed that electrons ran along circle tracks around the
nucleus. A. Sommerfeld (1868-1951) generalized these tracks and
assumed that also elliptic tracks were possible. He considered
relativistic effects, too. Everything was in good harmony with the
then experimental data (see e.g. \cite{somm}).

It is necessary to introduce that the models of Bohr and
Sommerfeld represented a partial success only. The theory
described well light frequencies but it was not able to predict
corresponding intensities. It was not possible, either, to apply
Bohr's postulates to atoms with a greater number of electrons.

Some questions remained unanswered: a cause why in principle
classical tracks were limited to some special orbits only. And
further, how the electrons pass from one orbit to the other one.
These transitions were spontaneous and indeterministic.

Various attempts were done to explain these characteristics in the
framework of classical physics. E.g., in 1924 Bohr, Krammers and
Slater assumed that the classical physics held also in atom
region, but the energy and momentum were not conserved
\cite{kram}. The idea was refused experimentally by Bothe and
Geiger \cite{bothe} and the conservation laws were confirmed in
1925 by A. Compton \cite{compt}.

   \subsection{Asymmetric world  }
 \label{ssecIII4}

In spite of the just mentioned problems the interpretation of
matter objects seemed to be clear: atoms consisted of positively
charged nucleus and one or more electrons running around; all
matter world consisting of such atoms. Some greater problems were
linked with the light that exhibited a kind of duality. There were
two different characteristics: slit interference phenomena that were
explained by wave behavior, and photoeffect when the light was
 similar to particle objects.
Asymmetry existed now between these two worlds (matter and light)
that were earlier quite different. And the question arose: Is it
possible to describe the light interference on the basis of
particle characteristics? or: Is it possible to describe both the
worlds on similar grounds?

Two different answers were given to these questions. First, W.
Duane \cite{duane} tried to show that interference phenomena may
be explained on particle basis as scattering of individual
photons. He assumed that the photons going through a grid obtained
perpendicular momentum that was equal to a multiple of $h/L$ where
$L$ was the grid constant (i.e., the distance between two slits).
It means that the directions of individual photons would be given
by perpendicular momenta $p_\perp = \frac{h}{L}, \frac{2h}{L},
..... $, corresponding to measured maximum values in interference
picture.

The other attempt was undertaken by L. de Broglie \cite{brogl} who
started in principle from particle aspect of individual objects
(similarly as Duane). He combined the idea of Planck with
relativistic idea. If a particle has rest mass $m_0$ its energy is
$m_0c^2$; it is then possible to ascribe to it according to Planck
the following frequency
             \[  \nu_0 \;=\;\frac{m_0c^2}{h}.  \]
As to the frequency corresponding to moving particle L. de Broglie
proposed formula
              \[  \lambda \;=\; \frac{h}{p}  \]
where $p$ was particle momentum.

The duality idea of L. de Broglie was used as basic idea of
further progress. He was convinced that particles represented 
full reality in the world. However, even if practically all energy
was concentrated in particle the wave was linked according to him
with each particle; his pilot wave represented a kind of reality,
too. The duality idea of L. de Broglie was then strongly supported
by interference characteristics found experimentally for electrons
passing through a crystal \cite{davi}.

    \subsection{Schr\"{o}dinger equation }
 \label{ssecIII5}

The idea of L. de Broglie was made use of by E. Schr\"{o}dinger
who was convinced that all the matter world consisted primarily of
waves. In the past many nature phenomena were described with the
help of various wave equations; e.g., sound or electromagnetic
waves. For a monochromatic wave with the frequency $\nu
\;=\;\frac{E}{h}\;$ Schr\"{o}dinger proposed the following
equation
 \[  i\hbar\frac{\partial\psi({\mathbf r},t)}{\partial t} \;=\;
      -\frac{\hbar^2}{2m_0}\Delta\psi({\mathbf r},t)\;+\;
                                U({\mathbf r}) \psi({\mathbf r},t) \]

He was fully successful as he was able to reproduce practically
all results obtained earlier with the help of Hamilton equations.
Individual physical quantities were now represented by operators
${\mathbf A}$ acting on corresponding $\psi$-functions; it was
possible to write
  \[  A_f(t) \;=\; \int dV \psi^*({\mathbf r},t)\, {\mathbf A}\, \psi({\mathbf r},t) \]
where it was integrated over the whole volume and $A_f(t)$
represented time dependence of individual physical quantities.

The solutions of Schr\"{o}dinger equation could be written in the
form
    \[ \psi({\mathbf r},t) \;=\; \int dE\; c(E)\;e^{-iEt/\hbar}\,\psi_E({\mathbf r}) \]
where complex function $c(E)$ (determined by initial conditions at
time $t=0$) represented physical characteristics of the given
system and functions $\psi_E({\mathbf r})$ were square-integrable
solutions (Hamiltonian eigenfunctions) of time-independent
Schr\"{o}dinger equation
    \[   H\psi_E({\mathbf r}) \;=\; E\psi_E({\mathbf r}) \,. \]
 $H$ was the Hamiltonian
        \[  H \;=\;  \frac{{\mathbf p}^2}{2m} \;+\; U({\mathbf r})  \]
used standardly in the Hamilton approach, where now ${\mathbf
p}\equiv\frac{\hbar}{i}\frac{\partial}{\partial{\mathbf x}}$.

It followed from the preceding that solutions of time-dependent
Schr\"{o}dinger equation might be expressed as linear
superpositions of eigenfunctions of the Hamiltonian. They could be
represented, therefore, by vectors of the Hilbert space spanned on
these eigenfunctions (see \cite{vonne}). And one should conclude
that in the case of stable objects the representation should be
equivalent to that in the classical phase space. However, the new
representation led to a new (and strange) physics, which will be
discussed in following sections to greater details.

   \subsection{Interpretation of wave function  }
 \label{ssecIII6}

According to the preceding the Schr\"{o}dinger equation did not
mean anything more than a new formal mathematical approach of
calculating physical quantities (without any physical
interpretation of the wave function). Such a view changed very
much when in 1926 M.~Born \cite{born} attributed the meaning of
probability density to the wave function. However, the decisive
change came when N.~Bohr \cite{bohr28} linked the wave function
with properties of individual (microscopic) particles. The
microscopic world lost its ontological nature as matter objects
were interpreted on the basis of characteristics following fully
from wave nature.

While for L. de Broglie the waves were closely linked to the
existence of corresponding matter objects, the situation changed
now. The quantum-mechanical microscopic world exhibited
paradoxical properties, being once of particle kind and once of
wave kind. Also the individual particles exhibited very strange
characteristics being distributed in the space with the density
determined by the absolute value of wave function in the given
space point. All matter objects as well as photons were handled,
however, in the same way.

A series of physicists were not satisfied with such a physical
picture of the microscopic world. However, they abandoned
gradually their critical points of view when they were unable to
propose a more suitable mathematical model. Thus, Bohr's
interpretation became practically the only theory of the
microworld, even if the quantum-mechanical world was very strange.
The philosophical conviction of the whole society contributed very
much to given conclusions.

The only man who remained always critical was A. Einstein. He was
denoting the quantum mechanics as incomplete. Even if he was not
able to convince physical community to accept his view his
objections and his controversy with N. Bohr were discussed during
the whole last century and indicated all the time that some points
were not satisfactorily answered. Some questions
were even denoted as forbidden in the quantum mechanics. The main
points of the mentioned controversy will be summarized in the next
section.

   \subsection{Einstein - Bohr controversy }
 \label{ssecIII7}

Einstein started to criticize Bohr's Copenhagen quantum mechanics
from the very beginning. However, his first arguments were not
sufficiently reasoned and remained without any success. The
important argument was published by A. Einstein together with two
co-authors \cite{ein} in 1935.

A Gedankenexperiment (called commonly EPR experiment) was proposed
that should have demonstrated the deficiency of
the quantum-mechanical model. It was argued that according to
orthodox quantum mechanics the measurements performed on two
microscopic particles emitted from one object in opposite
directions should have been mutually dependent even if the
distance between two measurement devices were very large. N.~Bohr
\cite{bohr} opposed strongly and succeeded in convincing physical
community that such a characteristic belonged to properties of the
microworld.

The question was whether the $\psi$ function at one time point described 
fully all properties of a microscopic particle or whether some other
("hidden") parameters should have been added.
 The physical community was influenced by the argument of
von Neumann \cite{vonne} that the corresponding extension was
excluded by the quantum-mechanical mathematical model.
 \footnote{Such a conclusion followed from the superposition principle
 and from the representation of time-dependent solutions of Schr\"{o}dinger
 equation in the Hilbert space spanned on one set of Hamiltonian
 eigenfunctions; the time dependence could not be truly expressed
 under such conditions. The given problem will be analyzed and
 corresponding solution will be presented in Part~\ref{secIV}. }
 Any attention was not devoted to arguments of Grete Herrmann
\cite{gher}, either,  who tried to show that the argument of von
Neumann was a circular proof.

The situation started to change partially only when D. Bohm
\cite{bohm} showed that a hidden parameter was contained
already in the Schr\"{o}dinger equation. Bohm proposed also a
modification of EPR experiments: instead of measuring the
positions of involved particles he proposed to emit a pair of
particles with opposite spins and to measure their spin
orientations; the idea used then practically in all experiments
having been performed later. On the basis of Bohm's finding L. de
Broglie and J. P. Vigier returned to the older idea of pilot wave
and started to develop the theory of empty waves, carrying the
information only, but not any energy (see, e.g., \cite{sel}). 
An additional quantum
potential depending on space arrangements of individual
experiments (and on initial conditions, too) played a role in such
an approach. Particles could move along some definite tracks in
such a theoretical alternative.

 \subsection{Bell's inequalities }
 \label{ssecIII8}

The decisive progress (as to Einstein-Bohr controversy) was done when J. Bell \cite{bell} showed
that one assumption involved in von Neumann's approach was in
contradiction to real situation in the nature; for details see,
e.g., \cite{sel}. He modified this assumption in agreement with reality, 
which enabled the existence of hidden parameters.
 He applied the new mathematical model to the EPR experiments in Bohm's
modification and argued that some inequalities
between measured values should be fulfilled in the theory of
hidden variables, while they did not hold commonly in the standard
quantum mechanics.

This fact initiated many physicists to look for suitable
experiments enabling to solve the controversy between Einstein and
Bohr on experimental basis. It was recommended to use the emission
of two photons with opposite spins and to measure the coincidence
output with the help of two polarizers under different angles. A
detailed review of the whole problem may be found in paper of
Clauser and Shimony \cite{clau}.

In the given experiment two photons with opposite spin
orientations are emitted in opposite directions and pass through
two polarizers:
   \[  a\; |\; \longleftarrow\; o\; \longrightarrow \;|\;b \;\; . \]
The orientation of transversal spins is randomly and uniformly distributed.
Coincidence transmission is then measured at different angles
$\alpha$ and $\beta$ of individual polarizers and corresponding
probabilities are established. Let us denote the probability that the
left photon is transmitted by $a_\alpha$ and similarly for the
right photon -- by $b_\beta$. When both the photons are fully
independent at the moment of measurement four different
probabilities should fulfill (according to Bell) the following
inequality
     \[ a_1b_1 \;+\; a_2b_1 \;+\; a_1b_2 \;-\; a_2b_2 \;<\;2 \; \]
where $a_1$ and $a_2$ are average transmission probabilities of
the left polarizer for two angles and $b_1$ and $b_2$ -- the same
for the right polarizer.

Such a condition should hold for the given combination of
any four (2x2) probabilities while in the case of quantum
mechanics it should be violated for some angle combinations. And
thus, the physics seemed to be in the position when it should be
possible to decide the controversy between Einstein and Bohr on
experimental basis. Any violation of the given condition should
prefer the standard quantum mechanics, while in the opposite case
the hidden-variable theory would be preferred.

  \subsection{Orthodox and ensemble interpretations of QM model }
 \label{ssecIII9}

While any $\psi({\mathbf r})$ function should represent the
properties of individual microscopic particles its form may be
experimentally derived with the help of statistical results in
experiments performed with many objects prepared in the same way.
I. e., the experimental arrangement corresponds to a wave function
while experimental results are given by a statistical distribution
of measured values. That was interpreted as a consequence of an
"absolute" chance in behavior of microscopic objects, the results being
limited only by the predictions following from the
quantum-mechanical model.

Such a situation seemed for some physicists to be untenable. They
wanted, therefore, to relate the given wave function not to an individual
particle but directly to the measured
statistical distribution. Consequently, two different
interpretations of the quantum mathematical model were discussed:
Copenhagen (or orthodox) and ensemble (or statistical); see, e.g.,
\cite{home}. In the latter case one might practically avoid the
mentioned paradoxes (following from the absolute chance in interactions of
individual microscopic objects). However, at the
same time one should conclude that the given mathematical model
corresponds to incomplete description of individual objects; and some
"hidden" variables should be added to describe the given situation
in a complete way.

The extented mathematical model corresponding practically to the mentioned 
ensemble interpetation will be described in Part~\ref{secIV}. 
%
 \subsection{Results of EPR experiments  }
 \label{ssecIII10}

The experiments initiated by Bell's inequalities were performed in
principle in 1971-82. Additional experiments performed yet later
did not bring anything new. The final results (see L. Aspect et
al. \cite{asp}) might be formulated in the following way:  \\
\hspace*{2mm} $-$ inequalities derived by Bell have been surely
 violated;    \\
\hspace*{2.15mm} $-$ experimental results have been found to be
practically in harmony with quantum-mechanical predictions.

In the next following years these results were often being denoted
as victory of quantum mechanics. However, gradually such voices
have decreased as it has become always more evident that any of
the quantum-mechanical problems has not been solved. And it was
necessary to go deeper into basic assumptions on which the whole
quantum-mechanical model was built up.

It has been possible to show that the refusal any hidden-variable alternative
based on the results of EPR experiments have been
strongly influenced by two additional unjustified statements: \\
 \hspace*{2mm} $-$ in his book F. Belifante \cite{belif} has stated that the
Malus law measured for light transfer through two polarizers and
being in harmony with quantum mechanics cannot be derived in the
framework of a hidden-variable theory; statement that  must be denoted as
false (see the next section);   \\
 \hspace*{2mm} $-$ it has been usually stated that the inequalities
derived by Bell have been valid for any hidden-variable
alternative as any assumptions have not been involved in their
derivation, which is not true, either; see Sec.~\ref{ssecIII12}.

Both these statements will be analyzed in the next two sections.
Hidden-variable model (theory) will be then presented in Part~\ref{secIV} that
should be surely preferred to the standard quantum mechanics.

 \subsection{Malus law for two polarizers }
 \label{ssecIII11}

As mentioned the quantum-mechanical interpretation of EPR
experiments seems to have been strongly supported by the argument
of Belifante who stated wrongly that the predictions of a hidden-variable
theory had to differ significantly from those of
quantum-mechanical model (see the corresponding graphs on p. 284
in Ref. \cite{belif}). According to a
hidden-variable theory the transmission of a photon through a
polarizer pair (or of two equally polarized photons in coincidence
arrangement) should equal
 \[ p_2(\alpha)\;=\; \int_{-\pi/2}^{\pi/2}p_1(\lambda)
         \;p_1(\lambda-\alpha)\;d\lambda  \]
where $p_1(\lambda)$ is transmission probability through one
polarizer; $\lambda$ - deviation of photon polarization from the
axis of the first polarizer; and $\alpha$ - the angle deviation of
 the second polarizer. The same formula    
holds in both the arrangements if the photon polarization does not
change in passing through a polarizer.

\begin{figure}[htb]
\begin{center}
\includegraphics*[scale=.4, angle= -90]{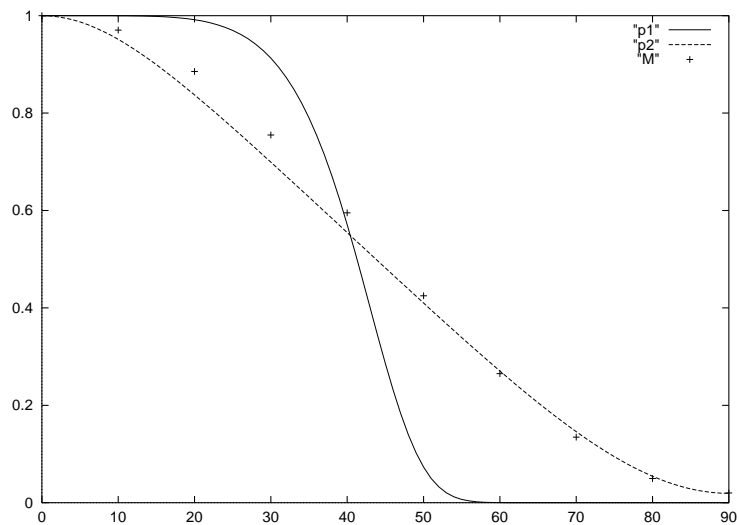}
   \caption{ {\it
Transmission probability through a polarizer pair leading to Malus
law;   $p_1(\lambda)$ - full line;  $p_2(\alpha)$ - dashed line;
             Malus law - individual points.   } } \label{fg1}
\end{center}
 \end{figure}
\vspace{0.2cm}

It has been known from the one-side experiments that it holds
 \[ p_2(\alpha) \;=\; M(\alpha) \;=\; (1-\varepsilon)\cos^2(\alpha)+\varepsilon  \]
where for real (imperfect) polarizers  $\varepsilon
> 0$ (generalized Malus law); $\varepsilon$ being very small.
Belifante has chosen
  \[    p_1(\lambda) \;=\; \cos^2(\lambda);  \]
i.e., he has interchanged quite arbitrarily the transmission
through one polarizer and through a pair of them. Malus law may be
easily reproduced in the hidden-variable alternative  if
$p_1(\lambda)$ is chosen  in a corresponding way as shown in Fig.
1 (full line); see \cite{loka2} for necessary details.

And one must conclude that any preference does not follow from
experimental data for the quantum-mechanical interpretation of
coincidence EPR experiments. Nothing prevents us, therefore, from interpreting
all available experiments on the basis of a hidden-variable
theory. This fact has opened also a new way to answer the question
concerning the time operator (or Pauli's) problem; see Part~\ref{secIV}.

  \subsection{Assumption in derivation of Bell's inequalities }
 \label{ssecIII12}

The way to paradoxical interpretation of quantum mechanics has
been paved (as already mentioned) by three mistakes: in addition
to von Neumann's original mistake two other mistakes have been
involved in corresponding argumentations.
  As the mistake of F.~Belifante has been explained in the
preceding section we will deal now mainly with the mistake
concerning Bell's inequalities.

Even if the mistake of von Neumann has been recognized in
principle by Bell its consequences have lasted practically until
now, having a close relation to (unphysical) interpretation of
general superposition principle.
 However, the problems have continued from another reason, too,
as the approach of von Neumann has been improved by J. Bell only
partially. A similar non-realistic (even if weaker) assumption has
been involved in derivation of Bell's inequalities, as well. They
were being derived in different ways (see, e.g., \cite{clau}), but
in all of them the same important assumption has been involved,
even if sometimes latently.

We will attempt now to explain the essence of this assumption,
necessary details being found in \cite{lok1}. The space
characteristics of a photon have not been described completely if
it has been characterized by its polarization $\lambda$ only. It
must be regarded (according to photoeffect) as a point-like (very
small) particle in any hidden-variable theory and characterized by
its position or by its impact parameter ${\mathbf f}$ when coming
to a polarizer. And one measures the average coincidence
probabilities of photon transfer
   \[  P_{\alpha, \beta} \;=\;
    \oint\!\! d\lambda \oint\!\!d{f}_\alpha
    \oint\!\!d{f}_\beta\;\; a_\alpha b_\beta \]
where $\oint$ represents corresponding averaging integral.

However, to obtain the given inequalities Bell had to make use of
the following condition (interchange between divers probability
pairs)
 \[ \oint\!\!d\lambda \oint\!\!d{f}_\alpha \oint\!\!d{f}_\beta\;\;
 a_\alpha b_\beta\,.\,a_{\alpha'} b_{\beta'}
  \;=\;\oint\!\!d\lambda\oint\!\!d{f}_\alpha\oint\!\!d{f}_{\beta'}
               \;\; a_\alpha b_\beta \,.\, a_{\alpha'} b_{\beta'} , \]
which destroys particle property of the photon making the
probabilities independent of the value of impact parameter. The
same assumption is involved in all kinds (shown in Ref.
\cite{clau}) of their derivations, even if in some of them it is
not explicitly mentioned (see \cite{lok1}).

It means that the measuring device has been regarded as a black
(or at least semi-black) box in a similar way as in the standard
quantum mechanics. Actual localization of a photon has been
omitted. And one must conclude that the violation of Bell
inequalities in coincidence polarization experiments is quite
 irrelevant as to the support for standard quantum mechanics.
 \footnote{In some papers the problem of the maximum expectation
value of Bell operator ${\mathbf B}$ (corresponding to the
combination of four measured quantities) discussed in
Sec.~\ref{ssecIII8} has been solved on the basis of the properties of
corresponding operators in the Hilbert space. It has been assumed
that the Hilbert space has consisted of the product of two
subspaces corresponding to individual measuring devices $a$ and
$b$; measurements of transmission probabilities being represented
by operators ${\mathbf a}$ and ${\mathbf b}$.
 The given problem has been discussed, e.g., in Ref. \cite{revz}. Under
different assumptions three numerical limit values have been
derived: $2$, $\;2.{\sqrt 2}$, and $\;2.{\sqrt 3}$. The first
value (Bell's limit) has been obtained if the commutativity of
operators ${\mathbf a}$ and ${\mathbf b}$ has been combined with
the additional assumption -- the commutativity of operators
${\mathbf a}$ or ${\mathbf b}$ inside individual subspaces --
equivalent to the case of Bell and discussed in Sec.~\ref{ssecIII12}.
The other value has been based on the mere mutual commutativity
between operators ${\mathbf a}$ and ${\mathbf b}$, which has
corresponded to the basic limit for any hidden-variable alternative
(being attributed mistakenly to the standard quantum mechanics in
\cite{revz}). The actual quantum-mechanical limit should
correspond to the case (denoted as unphysical in \cite{revz}) when
the only limitation has consisted in the requirement for measured
probabilities to lie in the interval $(0,1)$.  }

  \subsection{Time operator in Hilbert space  }
 \label{ssecIII13}

In the preceding sections we have discussed the problems related
to the controversy between Einstein and Bohr and to the EPR
experiments. And we have shown that there is not any argument
against Einstein's criticism. Now we will return to the problem of
time evolution of physical systems in the framework of the Hilbert
space and to the problem of defining time operator.

The evolution of a physical system is described by solutions of
time-dependent Schr\"{o}\-din\-ger equation that are then
represented in the Hilbert space spanned on one simple set of
Hamiltonian eigenfunctions. We have already mentioned that Pauli
showed in 1933 that the existence of the time operator fulfilling
condition
\begin{equation} \label{timop}
         i [H,T] \;=\; 1
\end{equation}
required for Hamiltonian spectrum to be continuous in the whole
interval $(-\infty,+\infty)$.

We have introduced, too, that there is a kind of contradiction
between the time-dependent Schr\"{o}dinger equation and
requirements following from the superposition principle in the
Hilbert space. It has been clear already for a longer time that the
solution cannot be reached in the framework of the given Hilbert
space and that the standard Hilbert space should be extended.
However, the way to the solution has been influenced by the
problem concerning the exponential phase operator, which will be
mentioned now.

The problem of time operator in quantum mechanics started to be
solved by P. Dirac \cite{dirac} for the case of linear harmonic
oscillator already in 1927. The oscillator has been described by
Hamiltonian
  \[  H \;=\; \frac{p^2}{2m} \;+\; kq^2 \; .  \]
 It has been possible to introduce operators
 \[ \mathcal{E}\;=\; (a a^\dag+1)^{1/2}a\:,
             \;\;\;  \mathcal{E}^\dag  \;=\; a^\dag (aa^\dag +1)^{1/2} \]
where annihilation and creation operators have been defined by
  \[ a \;=\; p - im\omega q, \;\;\; a^{\dag} \;=\;  p+im\omega q\;,
                              \;\;\;\; \omega=\sqrt{\frac{k}{m}} \; . \]
It holds
   \[  [H,\mathcal{E}] \;=\; -\omega \mathcal{E}\;, \;\;\;
          [H,\mathcal{E}^\dag] \;=\; +\omega \mathcal{E}^\dag \; ,
                          \;\;\mathcal{E}\mathcal{E}^+ = 1 \; ,  \]
and $\mathcal{E}$ corresponds to exponential phase operator
defined as
  \[ \mathcal{E}  = e^{-i\omega T}  .  \]

However, it has been shown later (1966) by Susskind and Glogower
\cite{sus}  that the operator  $\mathcal{E}$ is not unitary, but
only isometric:  $ \mathcal{E}^+\mathcal{E}  = 1 - |0><0|$, as it
holds
 \[ \mathcal{E}^\dag\mathcal{E}\:u_{1/2} \;\equiv\;  0 \;\neq\; u_{1/2}. \]
 It means that the unitarity condition is not fulfilled
for the state vector corresponding to the minimum-energy (vacuum)
state.

And the question has been whether this problem may be correlated
to the problem of Pauli or not. And how both the problems might be
removed: whether by one common extension of the Hilbert space or
whether they represent two different problems. The answers to
these questions will be given in Part~\ref{secIV}. Some additional
remarks will be introduced yet in the next section.

  \subsection{Quantum mechanics and microworld paradoxes }
 \label{ssecIII14}

Even if many physicists tried to contribute to the solution of the
mentioned quantum-mechanical problems their majority has seemed to
be quite content with the given situation. And we should ask what
the main reason was.

It was surely the philosophical attitude of the whole human
society. The positivistic philosophy convinced the scientists
already in the $19^{th}$ century that it is sufficient for the
physical science to represent measured values by mathematical
formulas only; to look for any deeper insight into the matter
world was even forbidden. Any metaphysical and ontological
considerations were being refused. On the other side some
(paradoxical) ideas following from contradictory mathematical
pictures of the world were being fully accepted (at least by some
important physicists).
 The belief in the quantum-mechanical paradoxes was then strongly
supported by far-east philosophies the ideas of which were being
widely spread in the western world mainly in the beginning of the
$20^{th}$ century.

Also the bivalent logic of Aristotle was being refused on the
basis of earlier conviction. The paradoxical picture of quantum
mechanics contributed to calling for many-valent logic. However,
there is not more any argument against it when the mentioned mistakes
have been removed. It is the full return of Aristotle's bivalent
logic into scientific (physical) approaches that might bring new
renaissance into the modern fundamental science.

The scientific method is represented maily by falsification
approach. According to it the reliable scientific conclusion may
be given by logical contradiction. Only the refusal of a
scientific hypothesis (or a set of such hypotheses) may represent
actual scientific truth. If no contradiction to experimental data
has been found the hypothesis may be denoted as plausible, but it
can be never regarded as verified. The scientific method consists
in the falsification, there is not any verification in the
science.

However, the idea of falsification was devaluated by the ideology
of falsifiability refusing any later parallel idea leading to the
same results even if it was not falsified and should have been
denoted as plausible. It is the falsification principle that
requires to look for parallel hypotheses, which opens new ways of
knowledge for the future. Forbidding competitive ideas makes
looking for truth practically impossible. The idea of
falsifiability (as used in physics in the last time) contradicts
the falsification method and should be denoted as anti-scientific.

 In addition to reasons following from the philosophical attitudes
there was also one physically-mathematical conviction (as already
mentioned) that contributed significantly to the fact that the problems 
of quantum mechanics remained unsolved. Practically all physicists trying to solve
the two problems (concerning time operator and exponential phase
operator) were convinced that both the mentioned deficiencies
represented one common problem, that should be solved at the same time,  
which was not, however, true.

In preceding sections we have analyzed main mistakes on which the
arguments supporting the standard quantum mechanics were based. It
has been shown that there is not any argument that would prefer
orthodox quantum mechanics to a hidden-variable theory in
 interpreting experimental data.
In the following Part~\ref{secIV} it will be shown how the individual problems
may be solved in two different ways; the agrrement of the extended mathematical
model to experimental data will be also discussed.

   \section{Extended quantum-theoretical mathematical model }
 \label{secIV}

  \subsection{Schr\"{o}dinger equation and superposition principle }
 \label{ssecIV1}

It is possible to say that the standard (orthodox) quantum
mechanics is based on two following basic assumptions:  \\
  \hspace*{3mm}  $-$ any state of a physical system consisting of
$N$ objects and its time evolution is represented by the wave
function $\psi(\{x_{k,j}\},t)$ that is obtained by the solution of
time-dependent Schr\"{o}dinger equation
 \[  i\hbar\frac{\partial}{\partial t}\psi(\{x_{k,j}\},t) \;=\; H\:\psi(\{x_{k,j}\},t) \]
where  $H$ is corresponding Hamiltonian
 \[ H \;=\; \Sigma^N_{j=1}\Sigma_{k=1}^3\frac{p^2_{k,j}}{2 m_j} \;+\;
                                 V(\{x_{k,j}\})    \]
and $\{x_{k,j}\}$ and $\{p_{k,j}\}$ are coordinate and momentum
components of individual objects; $V(\{x_{k,j}\})$ is the sum of
potential energies between $N$ individual mass objects;  it has
been put
 $ p_{k,j} \;=\; \frac{\hbar}{i}\frac{\partial}{\partial x_{k,j}} \;$
in the last equation ;  \\
  \hspace*{3mm}  $-$ any physical state is represented by a vector
in the Hilbert space being spanned on one set of Hamiltonian
eigenfunctions
   \[   H\: \psi_E(\{x_{k,j}\}) \;=\; E\: \psi_E(\{x_{k,j}\}) \]
and all states (vectors) are bound together with the help of
superposition principle.

Let us start with the problem concerning the application of
superposition principle in the region of physics. It is possible
to show that the general mathematical superposition principle
holding for all solutions of linear differential equations has
nothing to do with physical reality, as actual physical states and
their evolutions are uniquely defined by corresponding initial
conditions.
 \footnote{ It is being argued that any superposition of two solutions of
Schr\"{o}dinger equation is again a solution of the same equation.
However, such a statement is entitled in physics only if both
these solutions correspond to the same physical initial conditions
(characterized by different $\psi$-functions). Superposing
solutions belonging to divers initial conditions one obtains a
solution corresponding again to further fully different initial
conditions, which means that significantly divers physical states
have been combined in the framework of the standard
quantum-mechanical model
 in a physically unallowed way.}
They characterize individual solutions and represent properties of
a physical system, some of them being conserved during the whole
evolution. And consequently, any physical meaning cannot be
attributed to general superpositions in the Hilbert space (going
behind superposing the same physical trajectories).

Attributing physical interpretation to superpositions of different
solutions (belonging to divers physical states) introduces a drastic assumption into the physics without
any proof and any need. Statements that quantum mechanics
(including superposition rules) has been experimentally verified
must be regarded as wrong, too. All hitherto experimental tests
have concerned the time-dependent solutions of Schr\"{o}dinger
equation only (see also \cite{lok24}).

The other problem consists in the question whether one vector of
the given Hilbert space does represent a definite initial
condition. If one limits oneself to the physical system consisting
of two objects in CMS one can see easily that it is not so in the
standard Hilbert space. The wave function $\psi(\mathbf
r;0)$ determines initial vectors $\mathbf r_0$ and $\mathbf p_0$
(position and momentum). However, if the Hamiltonian is the
function of momentum squared it is evident from the
Schr\"{o}dinger equation that the time derivative of wave function
is the same for $\pm\mathbf p_0$. Consequently, two different
solutions evolving in opposite directions  cannot be distinguished
in the Hilbert space, in which the sign of the first time
derivative is not harmonized with that of $\mathbf p_0$. The Hilbert space
is to be extended and to consist of two different subspaces, the
approach being very similar to that proposed by Lax and Phillips
\cite{lax1} in the case of wave equation containing the second
time derivative of wave function; the problem being discussed in
Sec.~\ref{ssecIV2}.

In the preceding Part~\ref{secIII} we have described the problems
that accompanied quantum mechanics during the whole past century.
Now we should like to summarize their possible solutions. Only
main ideas and concepts will be explained in this paper;
corresponding mathematical structures have been described to greater
details already elsewhere (see Ref. \cite{last}).

 \subsection{Time operator and extended Hilbert space }
\label{ssecIV2}

In the standard quantum-mechanical model the Hilbert space has
been spanned on a simple vector basis consisting of Hamiltonian
eigenfunctions. In such a case the states corresponding to
opposite momenta (of different signs) have exhibited in principle
the same evolution (Hamiltonian being defined by momentum
squared). Consequently, it has not been possible to represent them
by clearly defined trajectories. And Pauli's critique \cite{pauli}
has concerned just this fact.

The given deficiency may be removed if the standard Hilbert space
is extended (doubled) in a way, as it was done by Lax and Phillips
 \cite{lax1} already in 1967 (see also \cite{lax2}) and derived
independently by Alda et. al. \cite{alda} on the basis of the
requirement for unstable particles to exhibit purely exponential
decay.

Let us demonstrate shortly the given Hilbert structure on the
example of a system consisting of two unbound particles. The
corresponding Hilbert space consists of two subspaces:
   \[   \mathcal{H}  \;\equiv\;  \{\Delta^- + \Delta^+\} ,  \]
each of them being spanned on one set of Hamiltonian
eigenfunctions. The individual solutions of Schr\"{o}dinger
equation are then represented by corresponding trajectories in the
total Hilbert space.

In the case of continuous Hamiltonian spectrum any point on such a
trajectory may be characterized also by expectation values of the
operator $R=\frac{1}{2}\{{\bf p}.{\bf q}\}$, where ${\bf q}$ and
${\bf p}$  are coordinates and momentum of one particle in CMS.
The states belonging to $\Delta^-$ are incoming states $(\langle
R\rangle <0)$, and those of $\Delta^+$ - outgoing states
(independently of the choice of coordinate system). The evolution
trajectory representing a solution of Schr\"{o}dinger equation
goes always in one direction from "in" to "out" and is
characterized by a set of initial conditions. There is a subset of
states $\psi_0$ characterized by
$\langle\psi_0|R|\psi_0\rangle=0$.

It is then possible to introduce also time operator T fulfilling
Eq. (\ref{timop}) and initial condition
$\langle\psi_0|T|\psi_0\rangle=0$.
 The subspace $\Delta^+$ ($\Delta^-$) corresponds then also to positive
(negative) expectation values of $T$. The structure of total
Hilbert space (corresponding to Schr\"{o}dinger equation) may be
defined by
  \[ \mathcal{H} \;=\; \overline{\bigcup_{t} U(t)\Delta^-}
                        \;=\;\overline{\bigcup_{t}  U(-t)\Delta^+}  \]
where
       \[     U(t) \;=\; e^{-iHt} \;\;  (t \ge 0)  \]
is evolution operator. It holds also, e.g.,
 \[ \Delta^+ \;\subset\; \overline{\bigcup_{t} U(t)\Delta^-} \;,\;\;
               \Delta^- \;\subset\; \overline{\bigcup_{t} U^\dag(t)\Delta^+} . \]

As in the case of two colliding particles the two different kinds
of states ("in" and "out") may be easily experimentally
distinguished it is useful to assume that $\Delta^+$ and
$\Delta^-$ are two mutually orthogonal subspaces. It is then also
possible to join an additional orthogonal subspace $\Theta$ that might
represent corresponding resonances formed in particle collisions
(see \cite{lax1}-\cite{alda}); i.e.
 \[ \mathcal{H} \;\equiv\; \{\Delta^-\oplus\Theta\oplus\Delta^+\}\;. \]
It is only necessary to define the action of evolution operator
between $\Theta$ and other subspaces in agreement with evolution
operators defined already in individual subspaces $\Delta^\pm$.

The evolution goes  in one direction, at least from the global
view; some transitions between internal states of $\Theta$ may be
reversible and chaotic. However, global trajectories tend always
in one direction;  see the scheme in Fig. 2.

In the case of discrete Hamiltonian spectrum (e.g., harmonic
oscillator) the wave function has similar $t$-dependent form.
However, the evolution is periodical as a rule. The Hilbert space
will consist of two subspaces, too (or rather of an infinite
series of such pairs if one wants to represent individual
subsequent periods); orthogonality between neighbor subspaces
being not required. The evolution may be again characterized by
trajectories corresponding to different initial conditions.
     \\

  \hspace*{8.0cm}   { $R \;=\; \frac{1}{2}\{\mathbf{p}.\mathbf{q}\};   \;\;
                                     \langle i[H,R]\rangle > 0$ }  \\
 \hspace*{3.34cm}  $\Delta^{(-)}\; \hspace{1.7cm} |
                                            \hspace{0.8cm} \;\Delta^{(+)}$ \\
 \hspace*{3.15cm}  { $\langle R\rangle <0 \hspace{1.4cm} |
                                      \hspace{0.8cm}\langle R\rangle >0$  \\
 \hspace*{3.095cm}  { $"in"$ } \hspace{0.1cm} $\searrow
                                   \hspace{0.4cm}     \longrightarrow  |
                               \hspace{0.8cm} { "out" } \longrightarrow$ \\
 \hspace*{3.05cm}  $ \_\_\_\_\_\_\_ \_\_\_\_\_\_\_\_\_\_\_\_\_\_|
            \_\_\_\_\_\_\_\_\_\_\_\_\_\_\_\_\_\_ $ \\
   \hspace*{4.5cm}$| \hspace{1.6cm}  \nearrow \hspace{1.00cm}  | $    \\
 \hspace*{4.5cm}$| \hspace{1.5cm} { \Theta}\hspace{1.40cm}| $      \\
 \hspace*{4.5cm}$|\_\_\_\_\_\_\_\_\_\_\_\_\_\_\_\_\_\_\_\_\_\_\_|    $ }      \\[1mm]
 Fig.2: {\it Scheme of the Hilbert space extended according to Lax
and Phillips (for a two-particle system); three mutually
orthogonal subspaces and direction of time evolution indicated
(continuous spectrum).}
\\

The expectation values of operator $R$ cannot be used now to
define individual points on given trajectories as they change
periodically (e.g., faster than $\sin\Phi$). It is, however,
possible to introduce time operator $T$, its zero expectation
value being identified with one zero of $\langle R\rangle$; and to
characterize
 individual points on a trajectory by its expectation values.
Different points may be characterized, of course, also by
expectation values of the phase operator $\Phi$; or (for one period) 
in analogy with the continuous case by those of
$\tan(\Phi/2)$ or $\cot(\Phi/2)$ increasing from $-\infty$ to
$+\infty$. For more details see \cite{last} (or also
\cite{lok24}).

The given extension (doubling) of the Hilbert space has enabled to
solve the problem of Pauli. Time operator $T$ is defined regularly
in the extended Hilbert space for both the kinds of Hamiltonian
spectra and may be expressed as a function of operator $R$, or of
operators ${\mathbf p}$ and ${\mathbf q}$. Each trajectory
represents the evolution of a physical system with given initial
conditions (at $t=0$). Only trajectories corresponding to the same initial
conditions may be superposed. The same physical evolution may be
pictured by a greater number of trajectories, which introduces a
new degree of freedom (or degeneracy), and we might ask whether it
does represent also some additional features of reality.

The given question is fully open at the present. It is, however,
possible to ask whether this degeneracy might be related to the
fact that the colliding particles are complex objects the internal
structures of which may be represented by further Hilbert
subspaces added in the form of direct products (in the framework
of corresponding hidden-variable theory).

The given extension of Hilbert space (according to Lax and Phillips)
does not solve, however, the unitarity problem of exponential
phase operator. It is possible, e.g., to define creation and
annihilation operators for harmonic oscillator in the same way and
with the same consequences as earlier. This problem may be solved
in a different way as proposed by Fain \cite{fajn} already in 1967,
too. His proposal remained misunderstood as physicists looked all
the time for the common solution with Pauli's problem. The
unitarity problem of exponential phase operator will be discussed 
in the next section (see also \cite{lok24,pk}).

  \subsection{Unitary exponential phase operator  }
\label{ssecIV3}

The unitarity problem of exponential phase operator has been
related practically to one special vector of the Hilbert space,
i.e., to the eigenvector corresponding to the minimal eigenvalue
of Hamiltonian (zero-energy vector). Fain showed that in the case
of linear harmonic oscillator the unitarity of the given operator
might be saved if the standard Hilbert space was doubled. Similar
solution was then discussed also by R. Newton \cite{newt} in 1980.
It is, of course, necessary to double the Hilbert space extended
already according to Lax and Phillips (or according to Alda et
al.); see the preceding section.

This total Hilbert space should consist of two identical mutually
orthogonal subspaces:
 \[  \mathcal{H} \;=\; \mathcal{H}_+ \oplus  \mathcal{H}_- \;   ,  \]
where $\mathcal{H}_+$ and  $\mathcal{H}_-$ are of Lax and Phillips
type and are distinguished with the help of special operator $J$
having two quantum numbers $(+1,-1)$. The orthogonality of the
subspaces remains conserved during the whole time evolution; it
holds (superselection rule)
 \[ U(t)\mathcal{H}_+ \subset\mathcal{H}_+,\;\; U(t)\mathcal{H}_- \subset\mathcal{H}_- . \]

The minimum-energy states in both the subspaces have been mutually
linked with the help of the annihilation-type operators. As shown
by Fain the states in the two mutually orthogonal subspaces (with
separated time evolution) may be distinguished by different signs
in the relation between the phase and the flowing time:
      \[\Phi \;=\;\pm \omega T \; ; \]
the sign being equivalent to the quantum values of the operator
$J$ (cp. \cite{bauer}).

The question is whether operator $J$ (the sign of the phase) may
be interpreted in a physical sense. It follows from the analysis
of three-dimensional oscillator \cite{pk} that it might be related
to the orientation of the co-ordinate system or to the orientation
of the corresponding component of resulting spin (or of angular
momentum of a two-particle system); see also \cite{last}.

   \subsection{Total Hilbert space and physical reality  }
\label{ssecIV4}

Both the problems concerning the regular description of time
evolution and the unitarity of exponential phase operator in the
framework of Hilbert space have been solved as two different
problems by suitable extensions of the standard Hilbert space. The
solution could hardly be reached in the past when the problems
were combined together or regarded as one common problem.

The extension proposed by Lax and Phillips is more important and
more basic. It might be made use of to represent truly
corresponding solutions of time-dependent Schr\"{o}dinger equation
for both the kinds of Hamiltonian spectra (continuous as well as
discrete). The operator $U(t)$ represents the evolution going in
one direction only; from "in" to "out" in the continuous case, and
alternatively from one subspace to another one in the discrete
periodical case. A point on individual evolution trajectory may be
characterized by expectation values of
 operator $T$; and/or by those of
operator $R$ (in continuous case)  or of phase operator $\Phi$
(resp. $tg(\Phi/2)$) in individual periods (in discrete case --
see Sec.~\ref{ssecIV2}). The given characteristics for the
discrete case have been demonstrated on the problem of
three-dimensional harmonic oscillator (see \cite{pk}) as the
simplified picture (in the linear case) does not provide a full
answer.

As to the continuous case it is useful to introduce the
orthogonality condition between "in" and "out" subspaces as
considered by Lax and Phillips. An additional subspace may be
included that may represent corresponding unstable resonances
formed in collision processes (see Sec.~\ref{ssecIV2}). 
The problem will be mentioned yet in Part~\ref{secV}.

To remove the non-unitarity of exponential phase operator the
further doubling of the Hilbert structure in addition to that
required by Pauli's problem has been necessary. This second
doubling differs from the preceding approach in that the evolution
in two subspaces remain now permanently separated and there is not
any mutual linkage by evolution operator, either; superselection
rule holding.
 The evolution inside both the individual subspaces is then quite
identical. However, the question should be put, whether the
quantum states belonging to different expectation values of
operator $J$ may become evident in some interactions of bound
systems of different kinds.

  \subsection{Hilbert space and classical phase space  }
\label{ssecIV5}

We have already mentioned that Schr\"{o}dinger and Hamilton
equations have described a physical system of stable objects in
the same way. Also the trajectories in the phase space and in the
extended Hilbert space (the first doubling) have exhibited
practically the same properties, describing intrinsically
irreversible evolution of the given physical system. In this sense
the new physical concept of reality is very similar to the case
when the extension is based on the rigged Hilbert spaces or
complex Hamiltonian eigenvalues (see e.g. \cite{abo7,abo}).
Collision processes and resonance formation (and their decay) may
be represented in principle in both the approaches in a similar
way.

However, the extension according to Lax and Phillips should be
preferred in the case of multi-channel collisions and decays.
E.g., inelastic collisions may be easily represented if additional
orthogonal "out"-spaces corresponding to different collision
channels are added; all states in individual "out"- and
"in"-subspaces being given by solutions of corresponding
Schr\"{o}dinger equations. It is only necessary to define
corresponding transition probabilities from an original
"in"-subspace to individual "out"-subspaces in agreement with
experimental data. The theoretical    predictions of such
probabilities should be then derived on the basis of concrete
collision mechanisms where internal structures of colliding
particles should be involved (elastic collisions, fireball
formation, stripping, and similarly).

It is also the multi-channel decay of an unstable particle that
may be easily included into such a scheme. Such a particle may be
represented by a special subspace orthogonal to all "in"- and
"out"-subspaces. Such a subspace should possess a greater number
of dimensions; the lowest possible dimension number being $N+1$
where $N$ is the number of decay channels (see  \cite{lok8}). The
given particle may be formed as a resonance or as a new particle
object produced in inelastic collisions.

The possibility has been now opened, too, to include more
systematically the influence of internal structures of
individual objects involved in different processes. They may 
be described by including
corresponding subspaces (representing stable or unstable
particles) in the form of tensor products; individual subspaces
being  mutually linked by corresponding evolution operators.

One may say that all research of microscopic objects starts from
collisions of two objects. The subspaces representing the
structure of individual objects may be more complex
(many-dimensional) or very simplified (one vector in extremum)
according to the goal of corresponding exploration; i.e., according to the
role being played by a given particle in final results. Combining
the approaches of Lax and Phillips (and of Fain) may provide a very
flexible and adaptable tool for describing and analyzing
individual experimental situations.

  \section{Concluding remarks  }
\label{secV}

According to the orthodox (Copenhagen) quantum-mechanical model
the microscopic physical reality should exhibit very strange
characteristics, differing rather fundamentally from the classical
picture of the world based on earlier (direct)
observations and measurements. The new picture has been accepted
by physical community even if any sufficient and reasonable
explanation has never been given how the observed properties of
macroscopic objects may arise from so divers and strange
properties of internal (microscopic) components.

Even if the criticism of Einstein was the subject of permanent
discussions in the past sufficient attention has not been ever
devoted to evident mistakes that have been periodically mentioned.
In the preceding we have discussed and analyzed the most important critical
points and mistakes and also corresponding solutions of theirs.
There is not more any fundamental difference between microscopic
and macroscopic world,  in contradistinction to what was believed
during the $20^{th}$ century.

Some main points of the new picture should be stressed:

 (i) The evolution of the matter world should be regarded as
continuous and irreversible change of space arrangements of
individual matter structures; including object changes during
collision and decay processes. Only special closed systems may
evolve periodically (however, not reversibly). Any application of
earlier uncertainty principle to individual matter objects represents
unacceptable hypothesis.

 (ii) The model based on the extended Hilbert space (with
narrowed superposition principle) corresponds fully to the
quantum-mechanical concept of W. Lamb \cite{lamb} who seems to
regard Schr\"{o}dinger equation as a natural extension of
classical physics to the microscopic world. The model is also in
agreement with the recently published results of U. Hoyer
 \cite{hoyer} who has argued that Schr\"{o}dinger equation may be
 derived from classical probability distributions of L. Boltzmann.

 (iii) The proposed mathematical model should be preferred not only on
the basis of theoretical arguments, but also on experimental
grounds. Experiments have been already performed the results of
which should be denoted in principle as falsification of the orthodox
quantum-mechanical model; see \cite{kra1,kra2,loka2}.

 (iv) The problem of stationary states of bound systems (e.g., atoms)
 must be newly solved; the Hamiltonian eigenvectors have
 not more any direct physical meaning, as they do not belong to
 time-dependent solutions of Schr\"{o}dinger equation. Other (deeper)
 reasons, e.g.,  for the existence of atom orbits and transitions
 between them, must be looked for.

 (v) As to the physical systems consisting of stable objects the
 use of extended Hilbert space leads to full agreement with classical
 picture. However, the inelastic collisions and the decay of
 unstable particle may be now described in the framework of the
 same mathematical structure. Also the influence of internal
 structures   of involved particles may be easily included into the
 given scheme; the degree of representation complexity being
 chosen according to experimental need.

 (vi) Superselection rules must be extended into the individual
 subspaces spanned on double set of Hamiltonian
 eigenfunctions. Only trajectories belonging to the same initial
 conditions may be superposed; otherwise, the physical picture
 might be strongly deformed in comparison with reality.

 (vii) The time operator $T$  as well as evolution operator $U(t)$
 may be defined in principle regularly in the whole Hilbert space
 (in all subspaces and for all physically reasoned transitions). An
 expectation value of operator $T$ may be attributed to any instant
 state of a given physical system. Such a state is then fully
 represented by one vector of the whole Hilbert space without
 exhibiting any strange (paradoxical) properties; ontological
 nature of the matter world being conserved.

And in the very conclusion: The main
milestones of the modern physics have been shown and discussed 
in the presented paper. We
have not gone into mathematical details, referring to
corresponding papers where necessary details may be found. The
same has concerned the newly proposed mathematical structure. The mathematical
details have been ommitted, attention being concentrated mainly to
corresponding physical ideas. They may be found in corresponding quoted papers
(see mainly Refs. \cite{last,pk}).

I should like to appreciate very much the help of Dr.
P.~Kundr\'{a}t who read carefully the whole manuscript and
contributed to improvement of the text.

{\small

\end{document}